\documentclass[aps,prd,superscriptaddress,nofootinbib,floatfix,preprintnumbers,12pt]{revtex4-2}

\pdfoutput=1
\usepackage{amsmath,amssymb,slashed,bm}
\usepackage{xcolor,multirow}
\usepackage{graphicx}

\usepackage{hyperref,dsfont}

\makeatletter
\g@addto@macro\bfseries{\boldmath}
\let\Hy@backout\@gobble
\makeatother

\begin{document}
\preprint{CERN-TH-2026-157}

\title{
A CKM blind spot:\\ probing $b$-column rescaling with kaons}

\author{Avital Dery}
\email{avital.dery@cern.ch}
\affiliation{CERN, Theoretical Physics Department, Geneva, Switzerland}

\begin{abstract}
    CKM unitarity triangle constraints are insensitive to New Physics scenarios in which the $b$-column is uniformly rescaled. This is because B physics data over-constrain the angles of the (bd) unitarity triangle, but only calibrate its normalization.  
    We identify this flat direction and derive model-independent constraints on the rescaling. 
    We find that $|\varepsilon_K|$ already constrains the rescaling at the same level as the direct measurement of the $b$-column normalization, with current data allowing for ${\cal O}(4\%)$ deviations. Projected inputs promote the kaon sector to be the leading probe.
    A toy model employing mixing with a vector-like up-type singlet provides a proof of principle for a possible UV realization.
\end{abstract}

\maketitle

\section{Introduction}
Increasing precision in kaon measurements and prospects for further kaon data allow us to probe scenarios about which we have so far been agnostic. 
While the picture arising from the B physics flavor program is beautifully consistent (up to a few curious anomalies), it remains to be seen whether or not constraints from another flavor sector fit into the same paradigm, dictated by the CKM mechanism.
As has been demonstrated in the literature (see, for example, Refs.~\cite{Chobanova:2017rkj,Dery:2021vql,DAmbrosio:2023irq,DAmbrosio:2024rxv,DAmbrosio:2025amb}), there exist viable New Physics (NP) models that discriminate between kaons and B mesons, and can significantly alter kaon observables without distorting B meson constraints. 
Here we argue that not only can NP be tucked away in yet-to-be-measured kaon parameters, but that deviations from the Standard Model (SM) could actually be hiding in B-sector parameters.

We first note that the well-known CKMfitter plot~\cite{Charles:2004jd} of constraints in the $(\bar\rho,\bar\eta)$ plane is almost entirely insensitive to an overall rescaling of the $b$-column of the CKM matrix,
\begin{equation}\label{eq:GenRescaling}
    V_{ib}^{\rm eff.} \, = \, V_{ib}\cdot a  \qquad  i=u,c,t \, .
\end{equation}
This follows from the fact that the  $(\bar\rho,\bar\eta)$ plane depicts the modulus and argument of the normalized CKM ratio~\cite{Wolfenstein:1983yz,Buras:1994ec},
\begin{equation}\label{eq:rhobaretabar}
    -\frac{V_{ud}V_{ub}^*}{V_{cd}V_{cb}^*} \equiv (\bar\rho + i\bar\eta) \, . 
\end{equation}
Therefore, the consistent picture of the overlaid B physics constraints on the CKMfitter plot would be unchanged by such a rescaling.
The insensitivity of the $(\bar\rho,\,\bar\eta)$ plane to the inputs $\lambda$ and $A$ reflects two directions the traditional fit does not probe: variations of $\lambda$, connected to first-row normalization and the Cabibbo anomaly, and a uniform rescaling of the $b$-column, under which Eq.~\eqref{eq:rhobaretabar} is exactly invariant and which is the subject of this work.

This blind spot can be phrased in terms of known constructions: the Universal Unitarity Triangle~\cite{Buras:2000dm}, built from $\sin 2\beta,\, \Delta m_d/\Delta m_s$ and $|V_{ub}|/|V_{cb}|$, consists exclusively of phases and ratios and is therefore exactly invariant under the rescaling of Eq.~\eqref{eq:GenRescaling}.
Normalization-type B-sector quantities, on the other hand, are all proportional to the product of the Wolfenstein parameter, $A$, and the scale factor, i.e.,
\begin{equation}
    V_{ub}^{\rm eff.},\, V_{cb}^{\rm eff.} \,\propto \, A\cdot a\, ,
\end{equation}
and similarly $\Delta m_d,\Delta m_s \propto (A\cdot a)^2$.
Since Wolfenstein $A$ and the scale factor $a$ appear only in this combination, B physics alone cannot disentangle their contributions.
Only quantities sensitive to the overall normalization of the $b$-column, requiring $|V_{tb}^{\rm eff.}|$ input, or a comparison with a second flavor sector, can probe $a$.
Precision kaon measurements have begun to provide exactly this information. As shown in Ref.~\cite{Dery:2025pcx}, the CKM sensitivity of precision kaon observables enters through a single rephasing-invariant combination, enabling CKM determinations independent of the B-sector. Contrasting the two determinations, which the SM CKM paradigm relates by unitarity, is the basis of the analysis below.

In the following, we first study the model-independent constraints on effective CKM column rescaling due to row and column normalization, as well as within the SM effective field theory (SMEFT), in Section~\ref{sec:mod-indep}. The heart of the analysis is presented in Section~\ref{sec:kaon}, in which we investigate to what extent this scenario is probed by existing and prospective kaon constraints. In Section~\ref{sec:VLQ}, we present an example NP realization, using a toy model where effective non-unitarity of the CKM matrix is achieved by mixing of the left-handed quarks with a vector-like ${\rm SU}(2)$ singlet state. We conclude in Section~\ref{sec:conclusions}.

\section{Constraints from CKM normalization and EFT}
\label{sec:mod-indep}

\subsection{Row and column normalization constraints}
\label{sec:row-column}
Measurements of CKM magnitudes constrain the scaling scenario of Eq.~\eqref{eq:GenRescaling}.
The strongest normalization constraint arises from the $b$-column, which satisfies the relation
\begin{equation}\label{eq:sumVibsqr}
    |V_{ub}^{\rm eff.}|^2+|V_{cb}^{\rm eff.}|^2 + |V_{tb}^{\rm eff.}|^2 = a^2\, .
\end{equation}
Using the PDG values for the measured elements, we arrive at
\begin{equation}\label{eq:upperbcolumn}
    -0.063\, \leq \, (1-a)_{\sum |V_{ib}^{\rm eff.}|^2}\leq \,0.044 \qquad \text{at } 2\sigma\, .
\end{equation}
This bound is controlled by the uncertainty on the measured $|V_{tb}^{\rm eff.}|$, currently at the level of $2.7\%$~\cite{ParticleDataGroup:2024cfk}. The determination is dominated by experimental systematics in the single-top cross-section measurement, with an irreducible theory/PDF component entering through the predicted cross-section~\cite{LHC_TopWG}.

From row unitarity, we have the relations
\begin{equation}
    |V_{ib}^{\rm eff.}|^2 = a^2(1-|V_{id}|^2-|V_{is}|^2)\, .
\end{equation}
Applied to the three rows, $i=u,c,t$, only the first row yields a meaningful constraint.
However, first row unitarity is in significant tension with current measurements, and scaling of the $b$-column does not help within existing bounds. Although in principle the effective CKM matrix of Eq.~\eqref{eq:GenRescaling} could have been useful for explaining the Cabibbo anomaly, we find that we would need $|1-a|\gtrsim 0.73$ in order to alleviate the tension significantly, in contradiction with the upper bound of Eq.~\eqref{eq:upperbcolumn}. This is driven by the smallness of $|V_{ub}^{\rm meas.}|^2$ compared to the first row deficit. Therefore the NP scenarios we study leave the Cabibbo anomaly untouched.

\subsection{EFT analysis}
\label{sec:smeft}
In the context of the SM effective field theory (SMEFT), NP is described by higher dimensional operators of the SM fields,
\begin{equation}
    {\cal L} = {\cal L}_{\rm SM} + \sum_{d>4}\sum_i \frac{C_i}{\Lambda^{d-4}}{\cal O}_i\, .
\end{equation}
At dimension six, the relevant operators in the Warsaw basis are~\cite{Grzadkowski:2010es},
\begin{align}
    \left({\cal O}_{\phi F}^{(1)}\right)_{ij} &= i(\phi^\dagger \overset{\leftrightarrow}{D}_\mu \phi)(\overline F_L^i \gamma^\mu F_L^j), \qquad  & \left({\cal O}_{\phi F}^{(3)}\right)_{ij} = i(\phi^\dagger \overset{\leftrightarrow}{D}_\mu^a \phi)(\overline F_L^i \sigma_a\gamma^\mu F_L^j), \\ \nonumber
    \left({\cal O}_{\phi f}\right)_{ij} &= i(\phi^\dagger \overset{\leftrightarrow}{D}_\mu \phi)(\overline f_R^i \gamma^\mu f_R^j)\, ,
\end{align}
where $F=Q,L$ and $f=u,d,\ell$.
The modifications to the $Z$ and $W$ boson couplings to fermions are given by
\begin{eqnarray}\label{eq:Deltag_smeft}
    (\Delta g_L^f)_{ij} &=& -\frac{1}{2}\left(C_{\phi F}^{(1)}\pm C_{\phi F}^{(3)}\right)_{ij}\frac{v^2}{\Lambda^2}, \qquad (\Delta g_R^f)_{ij} = -\frac{1}{2}(C_{\phi f})_{ij}\frac{v^2}{\Lambda^2}, \\ \nonumber
    (\Delta U_L^f)_{ij} & =& (C_{\phi F}^{(3)})_{ij}\frac{v^2}{\Lambda^2}\, .
\end{eqnarray}
In Eq.~\eqref{eq:Deltag_smeft}, $(\Delta g^f_{L,R})$ denote the corrections to the neutral current couplings of the left- and right-handed fermions to the $Z$ boson, where the plus (minus) sign in $\Delta g_L^f$ corresponds to fermions of weak isospin $-1/2 \,(+1/2)$. The deviation in the charged current couplings to the $W$ boson is given by $(\Delta U_L^f)$.
Scaling of the CKM $b$-column corresponds to turning on $(\Delta U_L^Q)_{ib}$ via the Wilson coefficients $(C_{\phi F}^{(3)})_{ib}$.

As is evident from Eq.~\eqref{eq:Deltag_smeft}, any modification of the charged current couplings unavoidably results in a modification of the neutral current couplings as well, since $\Delta U_L^f$ is induced by $C_{\phi F}^{(3)}$, which appears also in $\Delta g_L^f$.
Moreover, it can explicitly be seen that $\Delta g_L^f$ can vanish when $C_{\phi F}^{(1)} = \mp C_{\phi F}^{(3)}$, but not simultaneously for up-type and down-type quarks. 
We can formulate this statement via the sum rule,
\begin{equation}
    (\Delta U_L^f)_{ij} \, = \, (\Delta g_L^f)_{ij}^{-} - (\Delta g_L^f)_{ij}^{+}  \, ,
\end{equation}
i.e., a sizable modification to the CKM must appear either in the up-type, or the down-type couplings to the $Z$ boson (or both).
Current bounds impose~\cite{Efrati:2015eaa,ATLAS:2023eld,CMS:2021aly}, 
\begin{eqnarray}\label{eq:currentZbbtt}
    |(\Delta g_L)_{bb}| \, &<& \,  {\cal O}(10^{-3})\, ,  \\ \nonumber
    |(\Delta g_L)_{tt}| \, &\lesssim& \, 0.09 \qquad \text{at } \, 95\% \text{ C.L.}\, ,
\end{eqnarray}
for the flavor diagonal couplings, and~\cite{ATLAS:2023qzr}
\begin{eqnarray}\label{eq:currentFCNC}
    |(\Delta g_L)_{ut,ct}| \, &<& \,  {\cal O}(10^{-2}) \, , 
\end{eqnarray}
for the third generation up-type flavor off-diagonal couplings.
Given the nearly two orders of magnitude between the allowed deviations in the diagonal couplings to $b\bar b$ vs. $t\bar t$, in the next section we focus on models where the deviation in the down-type neutral current is structurally canceled, 
\begin{equation}
    (C_{\phi Q}^{(1)})_{bb} = -(C_{\phi Q}^{(3)})_{bb}\, ,
\end{equation}
and the neutral current deviation is pushed to the up-sector.
Then, a model inducing CKM $b$-column rescaling, as in Eq.~\eqref{eq:GenRescaling}, corresponds to,
\begin{equation}
    (\Delta U_L^Q)_{ib} = (a-1)\cdot V_{ib}\, ,
\end{equation}
resulting in 
\begin{equation}
    |(\Delta g_L^Q)_{i3}| = \left|(1-a)\right|\cdot |V_{ib}| = \begin{cases}
        |1-a|\cdot |V_{tb}| \qquad\quad \text{for } \, t t \\ 
        |1-a|\cdot {\cal O}(\lambda^2) \qquad \text{for } \, c t \\
        |1-a|\cdot {\cal O}(\lambda^3) \qquad \text{for } \, ut
    \end{cases}\, .
\end{equation}
The rough bounds of Eqs.~(\ref{eq:currentZbbtt},\ref{eq:currentFCNC}) then translate into the model-independent bound
\begin{equation}\label{eq:model-indep}
    |1-a|_{Z\bar t t} \lesssim 0.09\, .
\end{equation}
The correlation between charged-current (CKM) and $Z_{t\bar t}$ coupling deviations underlying Eq.~\eqref{eq:model-indep} was first derived in the context of models with exotic quarks in Ref.~\cite{delAguila:1998tp}. The analogous correlation for down-type singlets, with tree-level FCNC, was studied earlier in Ref.~\cite{Nir:1990yq}. Eq.~(10) expresses the same correlation model‑independently at dimension six.
The off-diagonal bound of Eq.~\eqref{eq:currentFCNC} translates
into the subleading constraint, $|1-a|\lesssim 0.2$.
A dedicated SMEFT global-fit analysis is beyond the scope of this work.

\section{Contrasting $b$-column rescaling with kaon constraints}
\label{sec:kaon}
The scenario of CKM $b$-column rescaling is constrained, model-independently, by kaon physics measurements. 
This is because kaon observables are sensitive to the fourth power of the Wolfenstein normalization parameter, $A$, and can be contrasted with B physics observables, which always measure $(A\cdot a)$. 

To be more concrete, kaon constraints are functions of a single complex rephasing-invariant CKM combination, identified in Ref.~\cite{Dery:2025pcx} as the natural complex variable for kaon CKM analyses (there denoted by $-R_{ct}\,e^{-i\theta_{ct}}$),
\begin{equation}\label{eq:vtsvtd}
    {\cal Z} \equiv\frac{V_{ts}^*V_{td}}{V_{cs}^*V_{cd}}= A^2\lambda^4((1-\bar\rho)-i\bar\eta)+{\cal O}(\lambda^6)\, .
\end{equation} 
The same rephasing invariant can be constructed out of B physics observables (together with $\lambda$, which we treat as precisely known and unaffected by the rescaling). 
One choice of basis for the CKM employs the following observable set,
\begin{equation}\label{eq:Rbgammaset}
    \Big\{|V_{ub}^{\rm eff.}|,\, |V_{cb}^{\rm eff.}|,\, \gamma,\, \lambda=|V_{us}|\Big\}\, .
\end{equation}
Since $\{|V_{cb}|,|V_{ub}|,\gamma,\lambda\}$ is a complete set of rephasing invariants, it can be used to construct the full unitary matrix.

In particular, kaon observables can be expressed via B physics inputs. The B-sector construction of the invariant takes the form (see Appendix~\ref{app:exp} for the derivation and an alternative construction),
\begin{equation}\label{eq:Bz}
    {\cal Z}^\text{B-phys.} (|V_{cb}^{\rm eff.}|,|V_{ub}^{\rm eff.}|,\gamma\,;\lambda)
    \approx \frac{|V_{cb}^{\rm eff.}|^2\lambda-|V_{cb}^{\rm eff.}V_{ub}^{\rm eff.}|e^{i\gamma}}{\lambda(1-\frac{1}{2}\lambda^2)}  =  \widetilde A^2\lambda^4((1-\bar\rho)-i\bar\eta) \, + \, {\cal O}(\lambda^6) \, ,
\end{equation}
where $\widetilde A \equiv A\cdot a$.
Kaon observables can thus be contrasted with the B-sector construction of the same invariant, lifting the flat direction.
We note that in the statistical analysis we do not truncate at ${\cal O}(\lambda^6)$ but rather use exact expressions constructed from the observables by building the unitary matrix.
The construction then satisfies (up to corrections of relative order $\lambda^4$),
\begin{equation}
    {\cal Z}^\text{B-phys.}(|V_{cb}^{\rm eff.}|,|V_{ub}^{\rm eff.}|,\gamma) \, = \, a^2 \cdot {\cal Z} \, .
\end{equation}

In the following we use the set of kaon observables,
\begin{equation}\label{eq:Kmeasurements}
    \Big\{|\varepsilon_K|,\quad {\cal B}(K^+\to\pi^+\nu\bar\nu),\quad {\cal B}(K_L\to\pi^0\nu\bar\nu) \text{ and } A_{\rm CP}(K\to\mu^+\mu^-)\Big\}\, ,
\end{equation}
schematically proportional to the combinations (see explicit expressions in Eqs.~(\ref{eq:epsKdep},\ref{eq:KplusB})),
\begin{equation}
    {\cal I}m({\cal Z}) \cdot {\cal R}e({\cal Z}),\quad {\cal I}m({\cal Z})^2 + {\cal R}e({\cal Z})^2  ,\quad {\cal I}m({\cal Z})^2\, ,
\end{equation}
with current uncertainties
\begin{equation}
    \Big\{4.6\%,\,\, 20\%,\,\, \text{NA}\Big\}\, ,
\end{equation}
where the third set have not been measured.
The two CP-violating modes share a single CKM dependence~\cite{Brod:2021hsj,Dery:2021mct,Brod:2022khx}, ${\cal B}(K_L\to\pi^0\nu\bar\nu) \propto A_{\rm CP}(K\to\mu^+\mu^-) \propto ({\cal I}m{\cal Z})^2$.
With projected uncertainties of ${\cal O}(20\%-30\%)$~\cite{KOTO:2025gvq,DAmbrosio:2025mxa}, these future measurements would not compete with $|\varepsilon_K|$ and ${\cal B}(K^+\to\pi^+\nu\bar\nu)$ in the one parameter rescaling scenario, and we do not include them in the fit. 
Their significance lies elsewhere: they complete the functionally independent set of precision kaon CKM observables, turning the kaon sector into an over-constrained CKM determination of its own~\cite{Dery:2025pcx}.
The analysis below therefore focuses on the two measured modes --- $|\varepsilon_K|$, which sets the current bound, and ${\cal B}(K^+\to\pi^+\nu\bar\nu)$, which drives future scenarios.

An important note on uncertainties: since kaon observables scale as $({\cal Z})^2$, they are sensitive to $a$ to the fourth power. A branching ratio measurement at the $5\%$ level, for example, then translates into an ${\cal O}(1.25\%)$ sensitivity to the scale factor, $a$.
Eq.~\eqref{eq:Bz} also shows that the uncertainties in the B physics determinations of the elements, $|V_{cb}^{\rm eff.}|,\, |V_{ub}^{\rm eff.}|$, directly affect the ability to constrain $a$. The current ${\cal O}(6\%-8\%)$ B physics uncertainty on ${\cal Z}^\text{B-phys.}$ translate into ${\cal O}(3\%-4\%)$ uncertainty on $a$, and currently dominate the bound set by $|\varepsilon_K|$.

\begin{table}[t]
\centering
\begin{tabular}{|l|c|c|c|}
\hline
Quantity & Value & $\,\,$ current $\sigma\,\,$ & $\,\,$ projected $\sigma\,\,$  \\
\colrule
\multicolumn{4}{|l|}{\textit{$B$-sector / CKM}}\\[2pt]
\hline
$|V_{cb}^{\rm eff.}|$            & $41.1\times10^{-3}$ & $3\%$    & $1\%$~\cite{Belle-II:2018jsg}         \\
$|V_{ub}^{\rm eff.}|$            & $3.82\times10^{-3}$ & $5\%$    & $2\%$~\cite{Belle-II:2018jsg}         \\
$|V_{tb}^{\rm eff.}|$  & $1.010$             & $2.7\%$  & $1.5\%$~\cite{Azzi:2019yne}  \\
$\sin\gamma$             & $0.91$        & $2.4\%$ & $0.6\%$~\cite{LHCb:2018roe}       \\ 
$\sin2\beta$         & $0.709$             & $1.6\%$  & $1\%$~\cite{Belle-II:2018jsg}          \\
$|V_{ts}V_{tb}^{\rm eff.}|$\,(from $\Delta m_s$) & $41.5\times10^{-3}$ & $2.2\%$ & $1\%$~
\cite{USQCD:2022mmc}\\
$|V_{td}/V_{ts}|$\,(from $\Delta m_d/\Delta m_s$) & $0.207$ & $1.4\%$ & $0.5\%$~
\cite{USQCD:2022mmc}  \\
$\lambda=|V_{us}|$   & $0.2250$            & $0.3\%$  & $0.3\%$           \\ 
\hline
\multicolumn{4}{|l|}{\textit{Kaon observables}}\\[2pt]
\hline
$|\varepsilon_K|$    & $2.228\times10^{-3}$ & $4.6\%$~\cite{Brod:2021hsj} & $3\%$ (see text)         \\
$\mathcal{B}(K^+\!\to\pi^+\nu\bar\nu)$ & $9.6\times10^{-11}$ & $20\%$~\cite{Chang:2026vvx} & $10\%,\,5\%$~\cite{Aebischer:2025mwl}  \\
\colrule
\end{tabular}
\caption{Inputs to the analysis: central values and uncertainties used in the current-data fit (taken from the PDG, Ref.~\cite{ParticleDataGroup:2024cfk}, unless otherwise stated), and the projected uncertainties assumed for the
future scenarios. 
Correlations among inputs are neglected throughout.}
\label{tab:inputs}
\end{table}

We construct $\chi^2$ functions with the parameters 
$\{|V_{cb}|,\,|V_{ub}|,\,\gamma,\,(1-a)\}$, using current figures for the kaon observables of Eq.~\eqref{eq:Kmeasurements}, together with the B-sector observables listed in Table~\ref{tab:inputs}. 
We then profile over $\{|V_{cb}|,\, |V_{ub}|,\,\gamma\}$, evaluate the one-dimensional $\Delta\chi^2(1-a)$ and extract the resulting $2\sigma$ bounds on $(1-a)$. 
We first present current and projected constraints from $|\varepsilon_K|$, then turn to ${\cal B}(K^+\to\pi^+\nu\bar\nu)$ and future projections.
%
\subsection{Current and future constraints from $|\varepsilon_K|$}
\begin{figure*}[t]
      \centering
      \includegraphics[width=1.\linewidth]{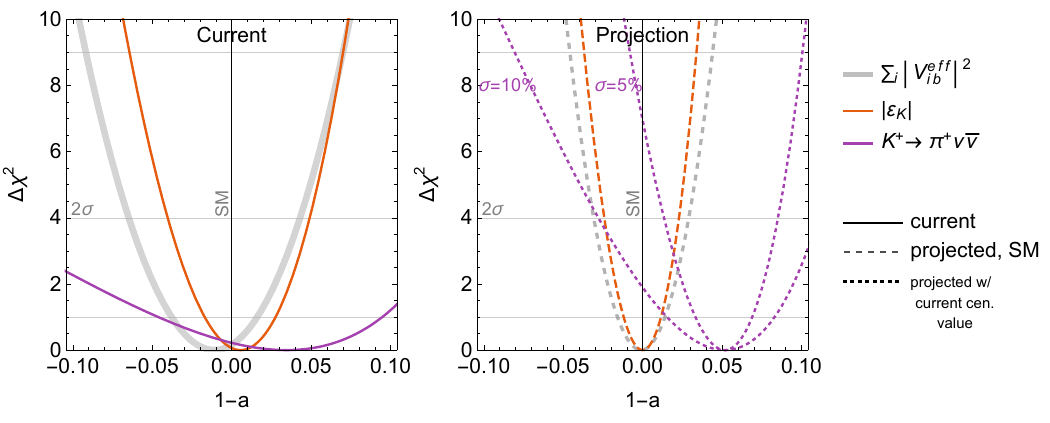}
     \caption{Values of $\Delta\chi^2$ as a function of the deviation in the scaling of the CKM $b$-column, $(1-a)$. In the left panel, the solid gray, orange and purple curves show the current bounds from $\sum_i|V_{ib}|^2$ (Eq.~\eqref{eq:upperbcolumn}), $|\varepsilon_K|$, and ${\cal B}(K^+\to\pi^+\nu\bar\nu)$, respectively.
     In the right panel, the dashed and dotted curves are future projections under the assumptions detailed in Table~\ref{tab:inputs}. Agreement with the SM is assumed for the $\sum_i |V_{ib}|^2$ and $|\varepsilon_K|$ projected curves, while scenarios where the current slight excess in ${\cal B}(K^+\to\pi^+\nu\bar\nu)$ persists are shown in dotted purple, for uncertainties of ${\cal O}(5\%,10\%)$.}
     \label{fig:kaonChiSqr}
    \end{figure*}

The measurement of $|\varepsilon_K|$ is sensitive to the following combination~\cite{Brod:2019rzc,Brod:2022har},
\begin{eqnarray}\label{eq:epsKdep}
    |\varepsilon_K| &\propto &  -{\cal I}m{\cal Z} \left({\cal R}e{\cal Z} - \bar\lambda\frac{\eta_{ut}{\cal S}_{ut}(x_c,x_t)}{\eta_{tt}{\cal S}_{tt}(x_c,x_t)}\right)\, ,
\end{eqnarray}
where ${\cal Z}$ is defined in Eq.~\eqref{eq:vtsvtd}, $\eta_{tt} = 0.55 (1\pm 4.2\%)$, $\eta_{ut}=0.402(1\pm1.3\%)$, ${\cal S}_{ut,tt}(x_c,x_t)$ are Inami-Lim functions~\cite{Brod:2019rzc,Inami:1980fz}, and $\bar\lambda\equiv \frac{\lambda}{|V_{cs}V_{cd}|} = 1+\frac{\lambda^2}{2}+{\cal O}(\lambda^4)$.
The $ut$ term is negative, ${\cal S}_{ut}\equiv {\cal S}(x_c)-{\cal S}(x_c,x_t) = -1.96\cdot 10^{-3}$~\cite{Brod:2019rzc}, so the charm contribution enhances $|\varepsilon_K|$ by $\sim 30\%$. 
Inserting the leading order form of ${\cal Z}$, from Eq.~\eqref{eq:Bz}, this becomes
\begin{eqnarray}\label{eq:epsKdepVcbVub}
    |\varepsilon_K| &\propto &  |V_{cb}V_{ub}|\sin\gamma \left(\lambda|V_{cb}|^2-|V_{cb}V_{ub}|\cos\gamma - \lambda\frac{\eta_{ut}{\cal S}_{ut}(x_c,x_t)}{\eta_{tt}{\cal S}_{tt}(x_c,x_t)}\right)\, .
\end{eqnarray}
We note that Eq.~\eqref{eq:epsKdepVcbVub} inherits the ${\cal O}(\lambda^2)$ relative precision of Eq.~\eqref{eq:Bz}, while the numerical analysis uses exact relations. 
We use the measured $|\varepsilon_K| = (2.228 \pm 0.105)\cdot 10^{-3}$~\cite{Brod:2021hsj} in the $\chi^2$ analysis, and find,
    \begin{equation}\label{eq:epsKconstraint}
     -0.040\,\leq\,(1-a)_{\varepsilon_K} \,\leq \, 0.049 \qquad \text{at } \, 2\sigma \, .
    \end{equation}
    The resulting $\Delta \chi^2$ is plotted in Fig.~\ref{fig:kaonChiSqr} (left), along with the analogous curve from the constraint on $\sum_i |V_{ib}|^2$, as in Eq.~\eqref{eq:sumVibsqr}.
    The $|\varepsilon_K|$ constraint is narrower than that from the direct normalization, but is centered at $(1-a)>0$. It therefore dominates the bound for $(1-a)<0$, while being mildly weaker for $(1-a)>0$.
    The displacement reflects the measured $|\varepsilon_K|$ sitting slightly above the SM prediction evaluated at Tab.~\ref{tab:inputs} inputs.
    In the case of $\sum_i |V_{ib}|^2$, the preference for $(1-a)<0$ is driven by the measured central value for $|V_{tb}^{\rm eff.}|$, which exceeds unity. 
    
    In terms of the CKM combination, $|\varepsilon_K|$ tests ${\cal I}m{\cal Z}\cdot {\cal R}e{\cal Z}$, whose B physics counterpart is approximately proportional to ${\cal I}m{\cal Z}^{\rm B-phys.}\cdot {\cal R}e{\cal Z}^{\rm B-phys.} \propto |V_{cb}^{\rm eff.}|^2|V_{ub}^{\rm eff.}|(\lambda|V_{cb}^{\rm eff.}|-|V_{ub}^{\rm eff.}|\cos\gamma)\sin\gamma$.
    The uncertainty dictating the sensitivity of Eq.~\eqref{eq:epsKconstraint} is dominated by the errors on $|V_{cb}|^3|V_{ub}|,\,|V_{cb}|^2|V_{ub}|^2$.
    Future reductions in these uncertainties would constrain the parameter space further. Once $\sigma(|V_{cb}|,|V_{ub}|)$ drop below $\sim 2\%$, the theory uncertainty on $|\varepsilon_K|$ becomes the limiting factor. For the future projections we use the projected reduced uncertainties as listed in Table~\ref{tab:inputs}.

    The non-parametric theory uncertainty on $|\varepsilon_K|$, currently at $4.6\%$~\cite{Brod:2022har}, is composed of a perturbative part ($3.0\%$, dominated by the scale variation of the NNLO short-distance factors $\eta_{tt}$ and $\eta_{ut}$) and a non-perturbative part ($3.5\%$, dominated by the long-distance parameter $\kappa_\varepsilon = 0.94(2)$ and the lattice bag parameter $\hat B_K$). 
    With an improved determination of the short-distance factors, anticipated by the authors of Ref.~\cite{Brod:2021hsj}, and continued lattice progress on $\hat B_K$ and the long-distance contributions~\cite{FlavourLatticeAveragingGroupFLAG:2024oxs,Bai:2023lkr}, a reduction to the $\sim 3\%$ level assumed in Table~\ref{tab:inputs} appears realistic on the timescale of the other projections.

    We find that, assuming a projected Gaussian error of $3\%$ on the determination of $|\varepsilon_K|$, and the reduced input uncertainties detailed in Table~\ref{tab:inputs}, the future sensitivity is at the level of
    \begin{equation}
        |1-a|_{|\varepsilon_K|}^{\rm Future} \, \lesssim  \, 0.02\qquad \text{at } 2\sigma \, .
    \end{equation}
    The projected $\Delta \chi^2$ curves are plotted in the right panel of Fig.~\ref{fig:kaonChiSqr}.

    %

\subsection{Future reach of ${\cal B}(K^+\to\pi^+\nu\bar\nu)$}
The branching ratio for the decay $K^+\to\pi^+\nu\bar\nu$ is being measured by the NA62 collaboration, proportional to~\cite{Buras:2015qea},
    \begin{eqnarray}\label{eq:KplusB}
        &\,&{\cal B}(K^+\to\pi^+\nu\bar\nu) \propto\left[({\cal I}m{\cal Z})^2 +\left({\cal R}e{\cal Z}+ \left(\frac{P_c\cdot \lambda^4}{X(x_t)}\right)\right)^2\right]\, ,
    \end{eqnarray}
where ${\cal Z}$ is defined in Eq.~\eqref{eq:vtsvtd}, $P_c = \left(0.404 \,\pm \, 0.024\right)\left(\frac{\lambda}{0.225}\right)^{-4}$, and $X(x_t) = 1.481 \pm 0.009$.
This mode is dominated by the real part, with a charm-induced offset. 
The current NA62 combined result~\cite{Chang:2026vvx},
\begin{equation}\label{eq:NA62measurement}
      {\cal B}(K^+\to\pi^+\nu\bar\nu) = (9.6^{+1.9}_{-1.8})\cdot 10^{-11}\, , 
\end{equation}
to be compared with the SM prediction, ${\cal B}(K^+\to\pi^+\nu\bar\nu)^{\rm SM} = (7.73\pm 0.61)\cdot 10^{-11}$, comes with an ${\cal O}(20\%)$ uncertainty, which translates into a $1\sigma$ uncertainty of ${\cal O}(5\%)$ on the extracted $(1-a)$.
In terms of $b$-column rescaling this measurement is currently subleading compared to $|\varepsilon_K|$ and would require an ${\cal O}(5\%)$ branching ratio uncertainty to significantly contribute.
It is interesting to note, that the current slight excess ($\sim 1\sigma$) compared to the SM prediction results in a preference for positive values of $(1-a)$, as shown in Fig.~\ref{fig:kaonChiSqr}, compatible with the prediction of the VLQ model of Sec.~\ref{sec:VLQ}.
If the current central value persists with more data, this would imply a preference for a value of $(1-a)\approx 0.05$. If, in addition, the branching ratio uncertainty were reduced to $10\%$,
within reach of the full NA62 dataset, the SM would be disfavored at the $1.4\sigma$ level; a further reduction to $5\%$ --- a HIKE-class experiment~\cite{Aebischer:2025mwl}, beyond the expected reach of NA62 --- would raise this to $2.6\sigma$. These hypothetical scenarios are shown in dotted purple in the right panel of Fig.~\ref{fig:kaonChiSqr}.

\section{A UV realization: mixing with vector-like singlets}
\label{sec:VLQ}
In the following we study specific realizations in terms of NP toy models, resulting in the CKM $b$-column rescaling of Eq.~\eqref{eq:GenRescaling}. We focus on mixing with vector-like (VL) ${\rm SU}(2)$ singlets, and study the leading existing constraints.
The aim of this section is to present a proof of principle that the scenario can be realized, rather than suggest a motivated NP theory. 

The relevant features of the toy models discussed are the following:
\begin{itemize}
    \item Mixing of the SM left-handed quarks with VL singlet fields results in mass eigenstates that are mixtures of ${\rm SU}(2)$ doublets and singlets, thus altering their coupling to the $W$ boson. Alignment with the $b$ quark mass eigenstate results in an effective scaling of the $b$-column of the CKM.

    \item The same singlet-doublet mixing also results in modified couplings to the $Z$ boson. As shown in Section~\ref{sec:smeft}, this is not specific to the VLQ models but a realization of the general dimension-six relation of Sec.~\ref{sec:smeft}.
\end{itemize}
The unavoidable effect on $Z$ couplings immediately rules out sizable rescaling via the most straightforward model, in which the $b$ quark mixes with a VL down-type singlet. Constraints on deviations in the $Zb\bar b$ coupling are at the order of $10^{-3}$.
Tree-level $Z$-mediated FCNCs and the associated violation of CKM unitarity in models with a down-type ${\rm SU}(2)$ singlet were analyzed in Ref.~\cite{Nir:1990yq}, including their implications for neutral-meson mixing and CP asymmetries.

We therefore focus on a model where an up-type VL field mixes with the left-handed \textit{up-type} quarks.
Consider the addition of vector-like quarks in the $\Psi_{L,R}\sim(3,1,2/3)\oplus(\bar 3,1,-2/3)$ representation.
The relevant lagrangian terms are then,
\begin{equation}\label{eq:lag}
    {\cal L}_{\Psi} \supset M_{\Psi}\bar\Psi_L \Psi_R + m_{\Psi u}\bar \Psi_L u_R+ Y_\Psi \bar Q_L \widetilde\phi \Psi_R \, .
\end{equation}
The mass terms for the up-type quarks now take the form
\begin{eqnarray}\label{eq:Mass}
    &\,&\frac{v}{\sqrt{2}}Y^u \overline{u_L}u_R + \frac{v}{\sqrt{2}}Y^\Psi u_L\Psi_R + m_{\Psi u}\overline\Psi_L u_R + M_{\Psi}\bar\Psi_L \Psi_R  \\ \nonumber
    &\,& = \begin{pmatrix}
        \overline u_L & \overline \Psi_L
    \end{pmatrix}\begin{pmatrix}
        \frac{v}{\sqrt{2}}Y^u & \frac{v}{\sqrt{2}}Y^\Psi \\ 
        m_{\Psi u} & M_\Psi
    \end{pmatrix}\begin{pmatrix}
        u_R \\ \Psi_R
    \end{pmatrix}\, ,
\end{eqnarray}
where $Y^u,\, Y^\Psi$ and $m_{\Psi u},\, M_\Psi$ are Yukawa and mass matrices in flavor space.
The lagrangian terms of Eq.~\eqref{eq:lag} result in left-handed up-type mass eigenstates that are mixtures of ${\rm SU}(2)$ doublets and singlets.

We consider a single vector-like quark aligned with the bottom quark mass eigenstate --- the unique choice yielding the uniform rescaling of Eq.~\eqref{eq:GenRescaling} --- i.e., mixing only with the up-type ${\rm SU}(2)$ partner of the bottom, $t_L^\prime \equiv  \sum_i V_{ib}^* {u_L}_i^{\rm mass}$. 
The couplings to the $W$ boson can be written in diagonal form,
\begin{eqnarray}\label{eq:Wint2}
    -\frac{g}{\sqrt{2}}\begin{pmatrix} \overline{u_L}^\prime & \overline{c_L}^\prime & \overline{t_L}^\prime \end{pmatrix}\slashed W^+\begin{pmatrix}
             d_L^{\rm mass} \\ s_L^{\rm mass}\\ b_L^{\rm mass}
         \end{pmatrix}\, ,
\end{eqnarray}
where the up-type quarks are written in the interaction basis.
Diagonalizing first the $2\times 2$ block of the mass matrix involving $t^\prime$ and $\Psi$, as in Eq.~\eqref{eq:Mass}, we can write,
\begin{equation}
  t_L^\prime \,=\, c_{\theta_b}\,    t_L
  + s_{\theta_b}\,\Psi_L^{\rm mass}\,,
\end{equation}
where $c_{\theta_b}\equiv\cos\theta_b$ and the angle $\theta_b$
parametrizes the bi-unitary diagonalization of the mass matrix. 
The charged current interaction of the light states then has the form
\begin{equation}
    -\frac{g}{\sqrt{2}}\begin{pmatrix} \overline{u_L}^\prime & \overline{c_L}^\prime & c_{\theta_b}\,\overline{t_L} \end{pmatrix}\slashed W^+\begin{pmatrix}
             d_L^{\rm mass} \\ s_L^{\rm mass}\\ b_L^{\rm mass}
         \end{pmatrix} = 
         -\frac{g}{\sqrt{2}}\begin{pmatrix} \overline{u_L}^\prime & \overline{c_L}^\prime & \overline{t_L} \end{pmatrix}\slashed W^+\begin{pmatrix}
             d_L^{\rm mass} \\ s_L^{\rm mass}\\ c_{\theta_b}\,b_L^{\rm mass}
         \end{pmatrix} \,
\end{equation}
Rotating to the up-type mass basis, we get an effective CKM matrix,
\begin{eqnarray}\label{eq:effCKM}
    (V_{\rm CKM}^{\rm eff.})_{ij} &=& (V_{\rm CKM})_{ij} \qquad \text{for }\, j\neq b\, , \\ \nonumber
    (V_{\rm CKM}^{\rm eff.})_{ib} &=& (V_{\rm CKM})_{ib}\cdot c_{\theta_b}\, ,
\end{eqnarray}
i.e.\ the entire $b$-column is rescaled by a common factor,
$c_{\theta_b}$.

The same singlet-doublet mixing modifies the neutral current,
\begin{equation}
  \mathcal{L}_{Z}\supset
  \frac{g}{c_W}\,Z_\mu
  \left[
    \bar{ t^\prime}_{L}\,\gamma^\mu\, t^\prime_{L}
    \Big(\tfrac12-\tfrac23 s_W^2\Big)
    +
    \bar{\Psi}_{L}\,\gamma^\mu\,\Psi_{L}
    \Big(-\tfrac23 s_W^2\Big)
  \right].
  \label{eq:Z-int}
\end{equation}
Rotating to the mass basis and projecting onto the light state, the coupling of $t_L^{\prime}$ is shifted,
\begin{equation}
  \Delta g^{t_L^\prime}\equiv g_L^{ t_L^\prime}-(g_L^{ t_L^\prime})^{\rm SM} =
  \left(\frac{1}{2}c_{\theta_b}^{2} - \frac{2}{3} s_W^2\right)
  -\left(\frac{1}{2}- \frac{2}{3} s_W^2\right)= -\frac{1}{2}s_{\theta_b}^{2}\,.
  \label{eq:dgZ}
\end{equation}
Re-expressing $t_L^\prime \approx \sum_i V^*_{ib} u_{iL}^{\rm mass}$ in terms of mass eigenstates, this shift induces the following deviations in the $Z$-couplings to the up sector,
\begin{equation}
  \Delta g^{(\hat u_i \hat u_j)}_{L}
  \;=\;
  -\frac{1}{2}s_{\theta_b}^{2}\,V_{ib}V^*_{jb}\,.
  \label{eq:upFCNC}
\end{equation}

This model exactly maps onto an EFT where $(C_{\phi Q}^{(1)})_{bb} = -(C_{\phi Q}^{(3)})_{bb}$ at the matching scale. The bound from deviations in the $Zt\bar t$ coupling is a bound on $\frac{1}{2}s_{\theta_b}^2 = (1-c_{\theta_b}) + {\cal O}(s_{\theta_b}^4)$, as in Eq.~\eqref{eq:currentZbbtt}.
We note that flavor-changing $Z_{sd},Z_{sb}$ couplings are regenerated at one loop. The former then contributes to $s\to d\nu\bar\nu$ via tree-level $Z$ exchange, while $|\varepsilon_K|$, requiring two insertions, is essentially unaffected. 

We note that misalignments away from the $t_L^\prime$ direction would modify this picture. Writing the single new mixing direction as $\hat n = \cos\varphi\, \hat t^\prime + \sin\varphi\, \hat n_\perp$, the effective CKM becomes $V^{\rm eff.} = \left[\mathds{1} - \left(1 - c_{\theta_b}\right) \hat n \hat n^\dagger\right] V\,,$ so that $b$-column rescaling is preserved up to corrections of relative order $s_{\theta_b}^2 \sin\varphi$. A misalignment component along the up-quark mass eigenstate is of particular interest, as it rescales the first row and could in principle accommodate the Cabibbo-anomaly deficit for $d_u \equiv (1-c_{\theta_b})\sin^2\varphi \approx 0.8\times 10^{-3}$, with the correct (deficit) sign guaranteed by unitarity of the enlarged matrix~\cite{Belfatto:2019swo,Belfatto:2021jhf,Branco:2021vhs}. The cross terms, however, open the $(bd)$ unitarity triangle into a quadrangle~\cite{Nir:1990yq}, with a fourth side of magnitude $2\sqrt{d_b\, d_u}\,|V_{ud}|$, where $d_b \equiv (1-c_{\theta_b})\cos^2\varphi \approx (1-a)$. For $d_b$ at the percent level and $d_u$ fixed by the Cabibbo deficit, this exceeds $|V^*_{ub}V_{ud}|$ and is excluded by the observed closure of the unitarity triangle, independently of the misalignment phase.
    Simultaneous realization of both rescalings at observable levels therefore requires two singlets, aligned with $\hat t'$ and $\hat u$ respectively, in which case both rescalings are exact and independent. Finally, since the $m_t^2$-enhanced contribution to $\Delta T$ scales with the top admixture, misalignment reduces it only in proportion to $d_b$ itself, and does not alleviate the oblique-correction tension at fixed $b$-column rescaling.

Beyond the model-independent constraints summarized in Section~\ref{sec:mod-indep}, this specific toy model induces deviations in electroweak precision observables and is constrained by collider searches. 
First, the VLQ heavy state is produced at the LHC via single and pair production, and is subject to collider bounds.
Current bounds on QCD pair production constrain the mass of the heavy state to be~\cite{ATLAS:2024gyc,CMS:2022fck},
\begin{equation}
    M_\Psi \,  \gtrsim \,  1.5 \, \text{TeV}\, .
\end{equation}
For mixing of roughly $1-c_{\theta_b}\approx \frac{1}{2}s_{\theta_b}^2 \approx |Y^\Psi|^2 v^2/(4M_\Psi^2)\sim 4\%$, this implies a Yukawa of order $|Y^\Psi|\sim 2.4$, in the perturbative range. Single production bounds are sensitive to the mixing angle directly, but result in subdominant bounds for $M_\Psi\gtrsim 1.5\,\text{TeV}$~\cite{ATLAS:2023pja,CMS:2024qdd}. 

Second, the model breaks custodial symmetry and generates a contribution to the oblique parameter $T$~\cite{Lavoura:1992np}, 
\begin{equation}
    \Delta T \approx \frac{3}{16\pi s^2_W c^2_W m_Z^2}\left[s_{\theta_b}^4 M_\Psi^2 + 2s_{\theta_b}^2 m_t^2\left(\log\frac{M_\Psi^2}{m_t^2}-1\right)\right]\, .
\end{equation}
At fixed mixing, the contribution grows with $M_\Psi$, and at the minimal allowed mass, $M_\Psi=1.5\,{\rm TeV}$, for a percent deviation, $\frac{1}{2}s_{\theta_b}^2\approx 1-c_{\theta_b} = 1\%$, the contribution already exceeds three times the uncertainty on the electroweak fit, $\sigma(T) \approx 0.06$~\cite{Alves:2023ufm}. 
A known way around this is to embed the VLQ in a custodially symmetric multiplet~\cite{Agashe:2006at}. 
Concretely, one completes the relevant fields into bi-doublets of ${\rm SU}(2)_L\times {\rm SU}(2)_R$, so that the new states fill out full custodial multiplets. This construction protects both the $T$ parameter and the $Zb\bar b$ coupling, at ${\cal O}(v^2/M_\Psi^2)$, while the CKM rescaling and corresponding deviation in up-type neutral current couplings remain.
A complete custodially invariant model, including the resulting spectrum and residual one-loop $S$ and $T$ contributions, is beyond the scope of this work.

We note that with $N$ up-type singlets the contributions add approximately linearly in the small-mixing limit, $\Delta T \approx \tfrac{3}{16\pi s_W^2 c_W^2 m_Z^2}\sum_i \left[ s_i^4 M_i^2
    + 2 s_i^2 m_t^2 \left(\log\tfrac{M_i^2}{m_t^2}-1\right)\right]$.
    At fixed total rescaling, $(1-a) \approx \tfrac{1}{2}\sum_i s_i^2$,
    distributing the mixing among $N$ equal states dilutes the
    $M_\Psi^2$-enhanced term by $1/N$, but not the $m_t^2$-enhanced term, which is fixed by the total rescaling and dominates at the minimal allowed mass: at $M_\Psi = 1.5$~TeV and $1-a = 1\%$, $\Delta T$ only decreases from $\approx 0.20$ to $\approx 0.16$ in the large-$N$ limit, and singlet contributions are positive-definite, precluding cancellations. Custodial protection therefore remains the structural solution.

The toy model thus demonstrates that the desired CKM structure can arise in a UV completion, although a realistic realization requires additional ingredients, such as custodial protection.
We note that the model above predicts $(1-a)>0$, since the scaling is controlled by a mixing angle, $c_{\theta_b}\leq 1$.
More generally, any scenario in which the rescaling arises solely from mixing with vector-like singlets gives $a \leq 1$, since the effective CKM is then a sub-matrix of a unitary matrix.
In principle, models with vector-like ${\rm SU}(2)$ triplets or additional charged vector bosons with couplings to the third generation can result in $(1-a)<0$. We do not study these further here.

\section{Conclusions}
\label{sec:conclusions}
The developing picture of precision kaon flavor measurements is invaluable, not only for the measurement of kaon parameters, but also for probing New Physics in the B-sector, otherwise obscured.
We identify a concrete blind spot of the standard CKM analysis: a uniform rescaling of the $b$-column, $V_{ib}^{\rm eff.} =a\cdot V_{ib}$.

We show that such a rescaling is nonetheless bounded, model-independently. Two different constraints dictate the current allowed range, $-0.040 \leq (1-a) \leq 0.044$, at $2\sigma$. The first, setting the upper bound, is the direct measurement of the $b$-column normalization, $\sum_i|V_{ib}^{\rm eff.}|^2$, resulting in $-0.063 \leq (1-a) \leq 0.044$, limited by the $\sim 2.7\%$ uncertainty on $|V_{tb}^{\rm eff.}|$ from single-top production. 
The second constraint, setting the lower bound, comes from the measurement of $|\varepsilon_K|$ --- a kaon observable that the rescaling does not affect --- dominated by the uncertainties on $|V_{cb}^{\rm eff.}|,|V_{ub}^{\rm eff.}|$ which enter through the comparison with the B-sector. This constraint yields $-0.040 \leq (1-a) \leq 0.049$.

These two leading probes are complementary, dominated by independent inputs on two different experimental tracks. The direct bound on the $b$-column normalization is pinned to $\sigma(|V_{tb}^{\rm eff.}|)$, whose single-top determination is theory/PDF and systematics limited. 
Kaon constraints, on the other hand, rely on contrasting $(V_{ts}^*V_{td})/(V_{cs}^*V_{cd})$ with B-sector combinations, subject therefore to $\sigma(|V_{cb}^{\rm eff.}|)$ and $\sigma(|V_{ub}^{\rm eff.}|)$, which are expected to significantly reduce with lattice QCD advancements, provided the tension between inclusive and exclusive determinations is resolved.
A significant improvement in B-sector determinations of $|V_{cb}^{\rm eff.}|,|V_{ub}^{\rm eff.}|$ and $\gamma$ will directly tighten kaon constraints on this CKM blind spot.
With projected B physics uncertainties at ${\cal O}(1\%)$, the $|\varepsilon_K|$ bound becomes theory-gated. Further progress then requires improvement in the theory inputs to the $|\varepsilon_K|$ prediction (the short-distance factors, $\eta_{tt},\eta_{ut}$, and the lattice inputs, $\hat B_K$ and $\kappa_\varepsilon$). Improving on the bound from $\sum_i |V_{ib}^{\rm eff.}|^2$, by contrast, requires progress on single-top systematics and on the PDF uncertainty of the predicted cross-section.

While $|\varepsilon_K|$ sets the bound, ${\cal B}(K^+\to\pi^+\nu\bar\nu)$ is the mode that could turn the blind spot into a signal.
The current NA62 central value sits $\sim 1\sigma$ above the SM prediction. If this excess persists as the uncertainty shrinks, it would develop into an intriguing hint of New Physics.
Beyond its magnitude, the sign of a future deviation would itself be informative, as it discriminates between UV scenarios.
Any model in which the rescaling arises solely from mixing with vector-like singlets predicts $(1-a)>0$, since the effective CKM is then a submatrix of a unitary matrix. 
Realizing $(1-a)<0$ requires new states with their own $W$ couplings: vector-like ${\rm SU}(2)$ triplets, or a NP vector sector. A percent-level measurement of $(1-a)<0$ would therefore disfavor the entire class of weakly coupled fermionic-mixing models, while $(1-a)>0$ --- the sign currently preferred by the ${\cal B}(K^+\to\pi^+\nu\bar\nu)$ central value --- is consistent with the minimal singlet scenario of Sec.~\ref{sec:VLQ}.

More broadly, this exercise demonstrates that contrasting precision CKM determinations across flavor sectors is a powerful null test of the Standard Model. With recent and expected advancements in precision kaon physics, we may still be able to probe sizable inconsistencies in the CKM --- the data could yet surprise us, and kaon physics could become a discovery channel once more.

\begin{acknowledgments}
    We gratefully thank Yossi Nir for useful discussions.
\end{acknowledgments}

\appendix

\section{Derivation of ${\cal Z}^{\rm B-phys.}$}
\label{app:exp}
We use the following expressions for B physics observables in terms of Wolfenstein parameters and the scale factor, $a$:
\begin{itemize}
    \item the angles,
    \begin{eqnarray}
     \sin 2\beta &=& \frac{{\rm Im}\Big[\left(-\frac{V_{cd}V_{cb}^*}{V_{td}V_{tb}^*}\right)^2\Big]}{\left|\frac{V_{cd}V_{cb}^*}{V_{td}V_{tb}^*}\right|^2} =
     \frac{2\bar\eta(1-\bar\rho)}{(1-\bar\rho)^2 +\bar\eta^2} + {\cal O}(\lambda^4)\, , \\ \nonumber
    \tan\gamma &=& \frac{{\rm Im}\left(-\frac{V_{ud}V_{ub}^*}{V_{cd}V_{cb}^*}\right)}{{\rm Re}\left(-\frac{V_{ud}V_{ub}^*}{V_{cd}V_{cb}^*}\right)} = \frac{\bar\eta}{\bar\rho} \, , 
    \end{eqnarray}

    \item and the magnitudes,
    \begin{eqnarray}
      |V_{cb}^{\rm eff.}| & =& A\lambda^2\cdot a + {\cal O}(\lambda^8)\, , \\ \nonumber
      |V_{ub}^{\rm eff.}| &= & A\lambda^3     \sqrt{\bar\rho^2+\bar\eta^2}\cdot a + {\cal O}(\lambda^5) \, , \\ \nonumber
      \Delta m_d &\propto & |V_{td}V_{tb}^{\rm eff.}|^2 = (A\lambda^3)^2 ((1-\bar\rho)^2+\bar\eta^2)\cdot a^2 + {\cal O}(\lambda^{10}) \, , \\ \nonumber
      \Delta m_s &\propto & |V_{ts}V_{tb}^{\rm eff.}|^2 = A^2\lambda^4\cdot a^2 + {\cal O}(\lambda^{6}) \, , \\ \nonumber
      \Delta m_d/\Delta m_s &\propto & |V_{td}/V_{ts}|^2 = \lambda^2(\bar\eta^2+(1-\bar\rho)^2)+{\cal O}(\lambda^4)\,.
    \end{eqnarray}
\end{itemize}

The above observables can be used to construct a basis for the entire CKM matrix.
For illustration, below are the approximate expressions for two choices of the CKM basis. 
First, using the complete set, $(|V_{cb}^{\rm eff.}|,|V_{ub}^{\rm eff.}|,\gamma,\lambda)$, we have
\begin{equation}
    \left[\frac{V_{ts}^*V_{td}}{V_{cs}^*V_{cd}}\right]^\text{B-phys.} \equiv {\cal Z}^\text{B-phys.} \approx \frac{|V_{cb}^{\rm eff.}|^2\lambda-|V_{cb}^{\rm eff.}V_{ub}^{\rm eff.}|e^{i\gamma}}{\lambda(1-\frac{1}{2}\lambda^2)} = \widetilde A^2\lambda^4((1-\bar\rho)-i\bar\eta) \, + \, {\cal O}(\lambda^6) \, ,
\end{equation}
with $\widetilde A\equiv A\cdot a$.

An alternative construction, using $(|V_{ts}V_{tb}^{\rm eff.}|,|V_{td}/V_{ts}|,\beta,\lambda)$ as a basis for the CKM is,
\begin{equation}
     {\cal Z}^\text{B-phys.} \approx\frac{|V_{ts}V_{tb}^{\rm eff.}|^2|V_{td}/V_{ts}|e^{-i\beta}}{\lambda(1-\frac{1}{2}\lambda^2)} = \widetilde A^2\lambda^4((1-\bar\rho)-i\bar\eta) \, + \, {\cal O}(\lambda^6) \, .
\end{equation}

\bibliographystyle{apsrev4-1}
\bibliography{biblio}

\end{document}